\documentclass[12pt]{article}
\usepackage{epsfig}
\usepackage{relsize}
\usepackage{graphicx}
\usepackage{pdproc}
\RequirePackage{xspace}
\newcommand{\nimBaseC}       {Nucl.\ Instr.\ and Methods\xspace}
\newcommand{\jprlBase}       {Phys.\ Rev.\ Lett.\xspace}
\newcommand{\jprBase}        {Phys.\ Rev.\xspace}
\newcommand{\jplBase}        {Phys.\ Lett.\xspace}
\newcommand{\jplb}       [1]  {\jplBase\ B~{\bf #1}}
\newcommand{\nim}       [1]  {\nimBaseC~{\bf #1}}
\newcommand{\jprl}      [1]  {\jprlBase\ {\bf #1}}
\newcommand{\jprd}      [1]  {\jprBase\ D~{\bf #1}}
\def\deltae{\ensuremath{\Delta E}\xspace}
\def\babar{\mbox{\slshape B\kern-0.1em{\scriptsize A}\kern-0.1em B\kern-0.1em{\scriptsize A\kern-0.2em R}}}
\def\CP{\ensuremath{C\!P}\xspace}

\def\to{\ensuremath{\rightarrow}\xspace}

\def\mes        {\mbox{$m_{\rm ES}$}\xspace}
\def\Bbar    {\kern 0.18em\overline{\kern -0.18em B}{}\xspace}

\def\CP                {\ensuremath{C\!P}\xspace}
\def\cp                {\ensuremath{C\!P}\xspace}
\def\to                 {\ensuremath{\rightarrow}\xspace}
\def\ra                 {\ensuremath{\rightarrow}\xspace}

\def\BR    {\ensuremath{\Gamma}}

  \makeatletter 
  \def\@cite#1{[#1]} 
  \makeatother    
\begin{document}

\newcommand{\BABARPubYear}    {04}
\newcommand{\BABARConfNumber} {154}
\newcommand{\SLACPubNumber}{10865}
\newcommand{\UCPubNumber}  {04-06}
\newcommand{\LANLNumber} {xxxxxxx}

\renewcommand{\thefootnote}{\alph{footnote}}

\begin{flushright}
\babar-PROC-\BABARPubYear/\BABARConfNumber \\
SLAC-PUB-\SLACPubNumber \\
UCHEP-\UCPubNumber \\
November 2004 \\
\end{flushright}

\title{MEASUREMENTS OF {\boldmath $B^-\to D^{(*)0}K^{(*)-}$} DECAYS RELATED
  TO {\boldmath $\gamma$}}

\author{GIAMPIERO MANCINELLI}

\address{Department of Physics, University of Cincinnati \\
ML 11, Cincinnati, OH 45211, USA\\
Representing the \babar\ Collaboration
}

\abstract{
We present measurements of branching fractions and \CP\ asymmetries of
several  $B^-\to D^{(*)0}K^{(*)-}$ decays, with the $D^{(*)0}$ decaying to
\CP-even, \CP-odd, and flavor eigenstates, that can constrain the \CP\ angle
$\gamma$ as well as the amplitude ratio $r_b=A(B\to u)/A(B\to c)$, using
methods proposed by Gronau, London and Wyler or Atwood, Dunietz
and Sony\cite{gronau}. We use data collected with the \babar\ detector at
the PEP-II asymmetric energy $e^+e^-$ collider at SLAC.
}

\normalsize\baselineskip=15pt

\section{Introduction}	

The unitarity of the Cabibbo-Kobayashi-Maskawa (CKM) matrix yields several 
relationships for its components, such as 
${V_{ub}^\ast}{V_{ud}}+{V_{cb}^\ast}{V_{cd}}+{V_{tb}^\ast}{V_{td}}=0$.
This describes the extent of \CP\ violation in the Standard Model (SM) in
the $B$ meson system and 
can be represented in the imaginary plane as a triangle, where the
angles ($\alpha$, $\beta$ and $\gamma$) can be written in terms
of the couplings between quarks:
\begin{equation}
\alpha \equiv {\rm arg} \left [-\frac{V_{td}V^*_{tb}}{V_{ud}V^*_{ub}}\right ] ~~
,~~ 
 \beta \equiv {\rm arg} \left [-\frac{V_{cd}V^*_{cb}}{V_{td}V^*_{tb}}\right ] ~~
,~~
\gamma \equiv {\rm arg} \left [-\frac{V_{ud}V^*_{ub}}{V_{cd}V^*_{cb}}\right ].  
\label{ckmangles}
\end{equation}
These angles can be extracted via \CP\ asymmetries 
measured in several decay modes of the $B$ meson. In particular 
$\gamma$ measurements can be made in modes which have
both $b\to c$ and $b\to u$ tree diagrams, which interfere.
The magnitude of the interference is determined by the
ratio of the two methods of decay. We report on
recent analyses which aim to measure the angle $\gamma$ with data
collected with the \babar\ 
detector\cite{detector}. All results are preliminary.
  
$B^-{\to}D^{(*)0}K^{(*)-}$ decays\footnote{Reference to the
charge-conjugate decays is implied  
throughout the text, unless otherwise stated.} can be used to constrain the
angle $\gamma$  of the 
Cabibbo-Kobayashi-Maskawa (CKM) matrix in a theoretically clean
way. 
The small branching fractions of these modes demand high efficiency and
the exploitation of as many decay modes as possible.
Two quantities are used to discriminate between signal and background: the
beam-energy-substituted mass 
$\mes \equiv \sqrt{(E_i^{*2}/2 + 
\mathbf{p}_i\cdot\mathbf{p}_B)^2/E_i^2-p_B^2}$
and the energy difference $\Delta E\equiv E^*_B-E_i^*/2$,
where the subscripts $i$ and $B$ refer to the initial 
{\ensuremath{e^+e^-}\xspace}\ system and the  $B$ candidate 
respectively, the asterisk denotes the CM frame, and the kaon mass 
hypothesis of the prompt track is used to calculate $\Delta E$.
The continuum background is also suppressed using topological
variables which exploit the fact that $B\overline B$ events are isotropic while
continuum events are jet-like.
Whenever we need to separate $B\ra D^{(*)0}K$ and $B\ra D^{(*)0}\pi$ events, 
we use measurements of the Cherenkov angle of the
prompt track\cite{detector}. At \babar\ 3 standard deviation separations
between the kaon and pion hypotheses  are
achieved for tracks up to 3.5 GeV/$c$. $\Delta E$  is also useful for this
purpose, as it depends on the mass assigned to the
tracks forming the $B$ candidate.
Backgrounds are characterized using simulation and off-resonance data.
The best candidate in each event is selected using observables which are
not used as 
inputs to the fits. Finally, unbinned
maximum likelihood fits are preformed to extract the signal yields.
The main systematic uncertainties are due to the characterization of the
probability density functions for signal and backgrounds and to the
particle identification, but many of these 
(e.g the absolute efficiencies) cancel when measuring ratios 
of branching fractions. For measurements of \cp\ asymmetries, possible
detector charge asymmetries (all consistent with zero) are also taken into
account.

\begin{figure}[htb]
\begin{minipage}{0.33\linewidth}
\begin{center}
\includegraphics*[width=1.\linewidth]{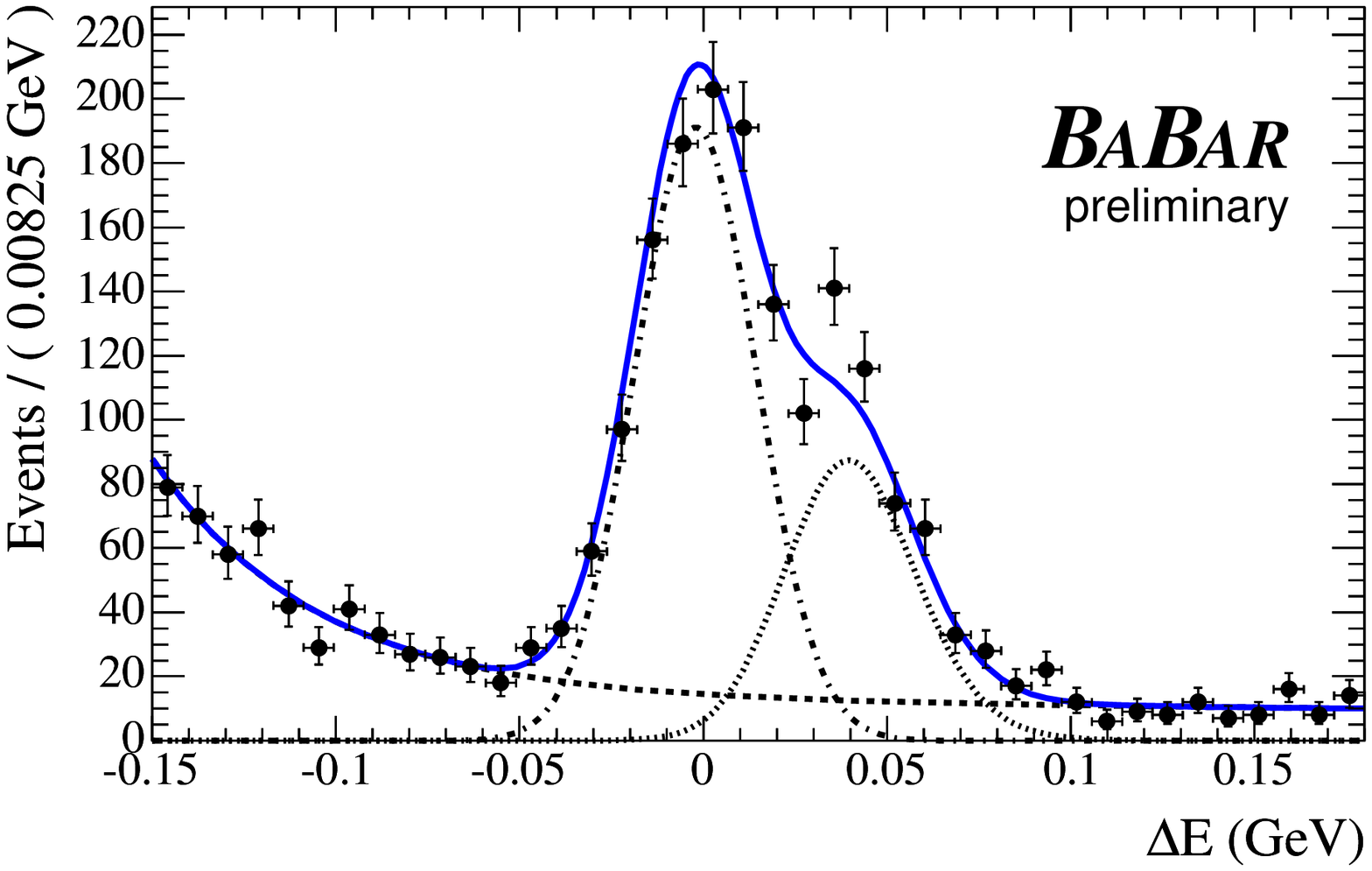}
\end{center}
\end{minipage}
\hskip-0.25cm
\begin{minipage}{0.33\linewidth}
\begin{center}
\includegraphics*[width=1.\linewidth]{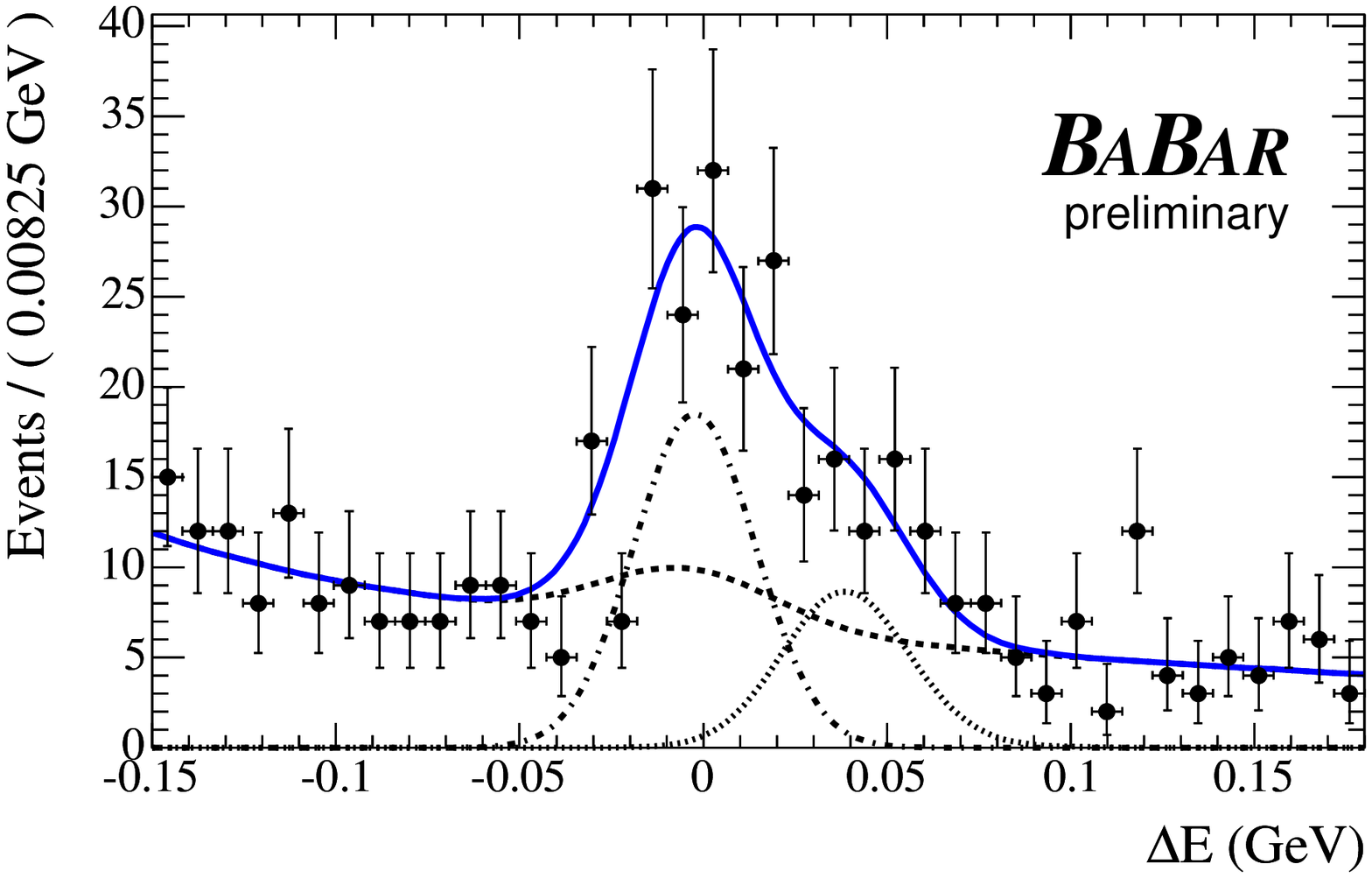}
\end{center}
\end{minipage}
\hskip-0.25cm
\begin{minipage}{0.33\linewidth}
\begin{center}
\includegraphics*[width=1.\linewidth]{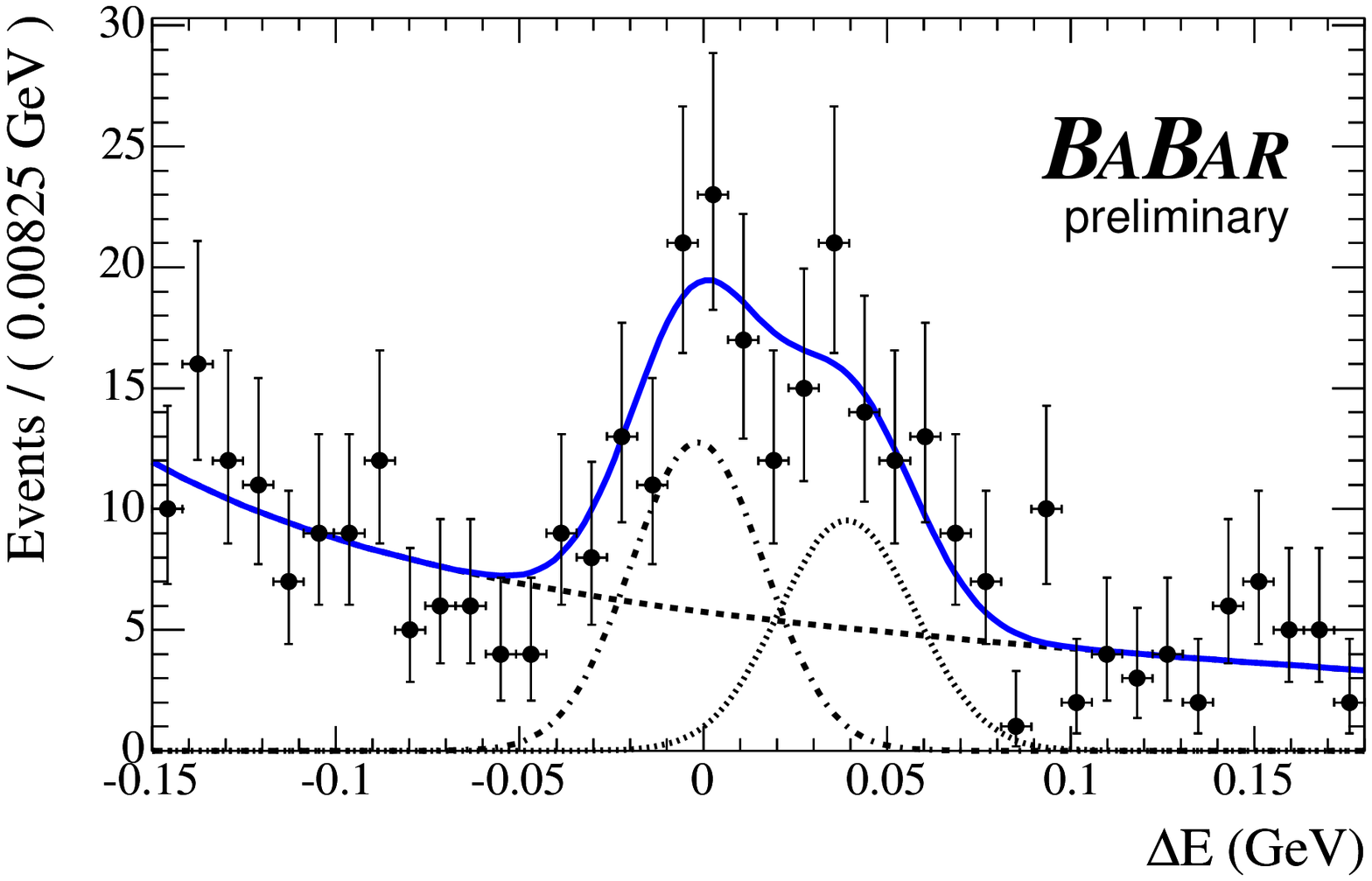}
\end{center}
\end{minipage}
\caption{$B^-{\ra}D^0K^-$ signal
after requiring that the prompt track be consistent with the
kaon hypothesis for the flavor (left), \CP$=+1$ (center), and \CP$=-1$
(right) eigenstates.   The
$B^{\pm} \to D^{0} \pi^{\pm}$ signal contribution on the right of each plot is shown as a
dotted line,  
the $B^{\pm} \to D^{0} K^{\pm}$ signal on the left as a dashed-dotted  line, and the background
as a dashed line. The total fit with all the contributions is
shown with a solid line.}
\label{8-0921_fig}
\end{figure}

\begin{table}[h]
\caption{\babar\ summary of results (GLW method). The first error is due to
statistical and the second to systematic uncertainties. The third, when
present, reflects possible interference effects in the final states with $\phi$ and $\omega$ resonances.}
{\begin{tabular}{|c|c|c|c|} 
\hline
Observables & $B^-{\ra}D^0K^-$& $B^-{\ra}D^{*0}K^-$&$B^-{\ra}D^0K^{*-}$ \\
\hline
\hline
&\\[-9pt]
$R_{\CP+}$ & $0.87\pm0.14\pm0.06$ & $1.09\pm0.26^{+0.010}_{-0.08}$ & $1.77\pm0.37\pm0.12$ \\
$A_{\CP+}$ & $0.40\pm0.15\pm0.08$ & $-0.02\pm0.24\pm0.05 $         & $-0.09\pm0.20\pm0.06$ \\
$R_{\CP-}$ & $0.80\pm0.14\pm0.08$ &                                & $0.76\pm0.29\pm0.06^{- 0.04}_{-\ 0.14}$ \\
$A_{\CP-}$ & $0.21\pm0.17\pm0.07$ &                                &
$-0.33\pm0.34\pm0.10$\\
& & &$(+0.15 \pm 0.10) \cdot (A_{\CP-} -A_{\CP+}) $ \\
\hline
\end{tabular}}
\label{table1}
\end{table}

\section{GLW related measurements}

In the SM, for $B^-{\to}D^{0}K^{-}$  decays, we have:
$R_{\CP\pm}/R_{{\rm non}-\CP}\simeq 1+r_b^2\pm2r_b\cos\delta_b \cos\gamma$
(in the absence of 
$D^{0}\overline{{D}^{0}}$ mixing) , where
\begin{equation}
R_{{\rm non}-\CP/\CP\pm}\equiv \frac{\BR(B^-\ra
D^0_{{\rm non}-\CP/\CP\pm}K^-)}{\BR(B^-\ra D^0_{{\rm non}-\CP/\CP\pm}\pi^-)},
\label{eq:rstar}
\end{equation}
$r_b$ is the ratio of the color suppressed $B^+\ra D^0K^+$ and color
allowed $B^-\ra D^0K^-$ amplitudes ($r_b \sim
0.1-0.3$), and $\delta_b$ is the
\cp-conserving strong phase difference between these
amplitudes. Furthermore, defining the direct \CP asymmetry
\begin{equation}
A_{\CP\pm}\equiv
\frac{\BR(B^-{\ra}D^0_{\CP\pm}K^-)-\BR(B^+{\ra}D^0_{\CP\pm}K^+) 
}{\BR(B^-{\ra}D^0_{\CP\pm}K^-)+\BR(B^+{\ra}D^0_{\CP\pm}K^+)}, 
\label{eq:cpa}
\end{equation}
we have:
$A_{\CP\pm}=\pm 2r_b\sin\delta_b\sin\gamma/(1+r_b^2\pm
2r_b\cos\delta_b\cos\gamma)$. Similar quantities and relationships exist for
the the modes $B^-{\to}D^{*0}K^{-}$ and $B^-{\to}D^{0}K^{*-}$, where the
corresponding $r_b$ and $\delta_b$ might have different values from the ones
for the $B^-{\to}D^{0}K^{-}$ mode. The unknowns $\delta_b$, $r_b$, and
$\gamma$ can be 
constrained from the measurements of $R_{{\rm non}-CP}$, $R_{\CP\pm}$, and
$A_{\CP\pm}$. The smaller
$r_b$ is, the more difficult is the measurement of $\gamma$ with this
method. 
\begin{figure}[htb]
\centerline{\psfig{file=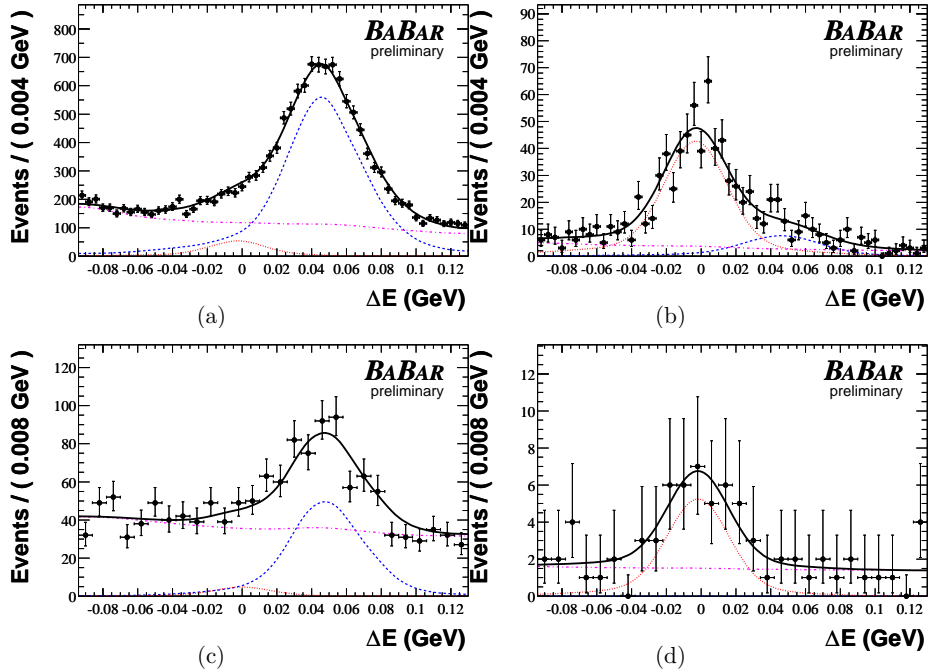,width=12.2cm}}
\vspace*{8pt}
\caption{Distributions of \deltae\ in the $B\rightarrow D^{*0}h$ sample, for
$D^0\rightarrow K^-\pi^+, K^-\pi^+\pi^0, K^-\pi^+\pi^+\pi^-$ ((a),
(b)) and  $D^0\rightarrow K^-K^+, \pi^-\pi^+$ ((c), (d)), before
((a), (c)) and after ((b), (d)) enhancing the $B\rightarrow
D^{*0}K$ component by requiring that the prompt track be consistent with the
kaon hypothesis and $\mes>5.27$GeV/$c^2$.  The
$B^{\pm} \to D^{*0} \pi^{\pm}$ signal contribution on the right of each
plot is shown as a 
dashed line,  
the $B^{\pm} \to D^{*0} K^{\pm}$ signal on the left as a dotted line, and
the background 
as a dashed-dotted line. The total fit with all the contributions is
shown with a thick solid line.}
\label{fit_con}
\end{figure}

\begin{figure} [hbt]
\centerline{\psfig{file=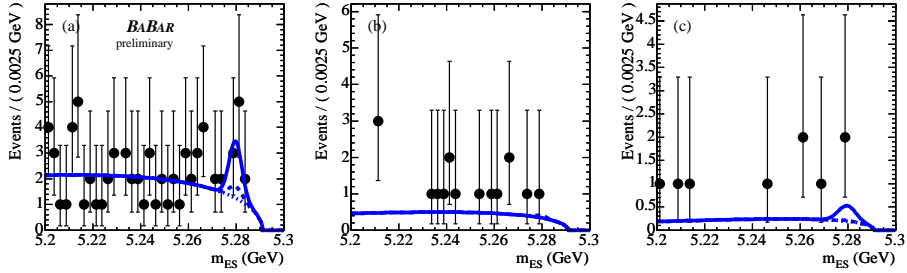,width=12.2cm}}
\vspace*{8pt}
\caption{Signal for the three suppressed decay modes used for the ADS
method: $B^{\mp} \to [K^{\pm} \pi^{\mp}]_D
K^{\mp}$ (a) ($4.7^{+4.0}_{-3.2}$ events), $B^{\mp} \to [K^{\pm}
\pi^{\mp}]_{D^*(D\pi)}$ (b)  ($-0.2^{+1.3}_{-0.8}$ events), and $B^{\mp}
\to [K^{\pm} \pi^{\mp}]_{D^*(D\gamma)}$ (c) ($1.2^{+2.1}_{-1.4}$ events).}
\label{figads}
\end{figure}

\begin{figure}[htb]
\begin{minipage}{0.42\linewidth}
\begin{center}
\includegraphics*[width=1.\linewidth]{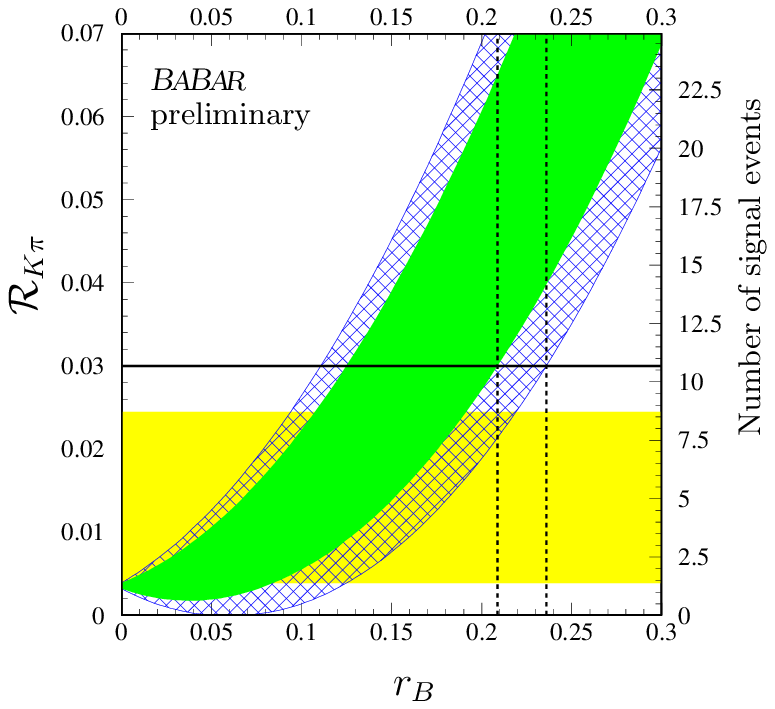}
\end{center}
\end{minipage}\hfill
\begin{minipage}{0.42\linewidth}
\begin{center}
\includegraphics*[width=1.\linewidth]{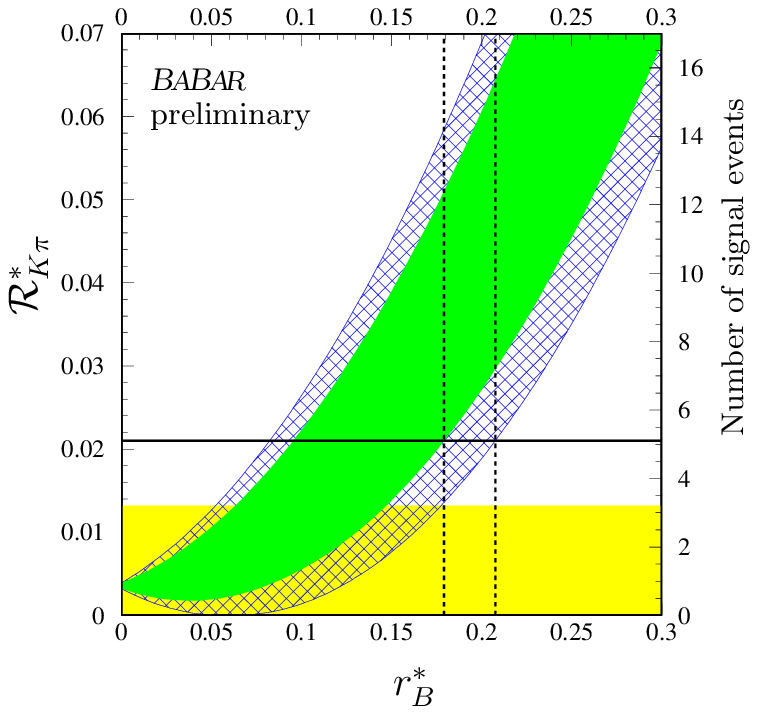}
\end{center}
\end{minipage}\hfill
\caption{Dependence of $R_{K\pi}$ on $r_b$ for the $D^{0}K$ mode (left)
and for the $D^{*0}K$ mode (right) using 
$0^o~<~\gamma,~\delta~<~180^o$ (hashed area) and the range
of $\gamma$ from CKM fits ($48^o~<~\gamma~<~73^o$).
}
\label{8-0948}
\end{figure}

\begin{figure} [hbt]
\begin{minipage}{0.42\linewidth}
\begin{center}
\includegraphics*[width=1.\linewidth]{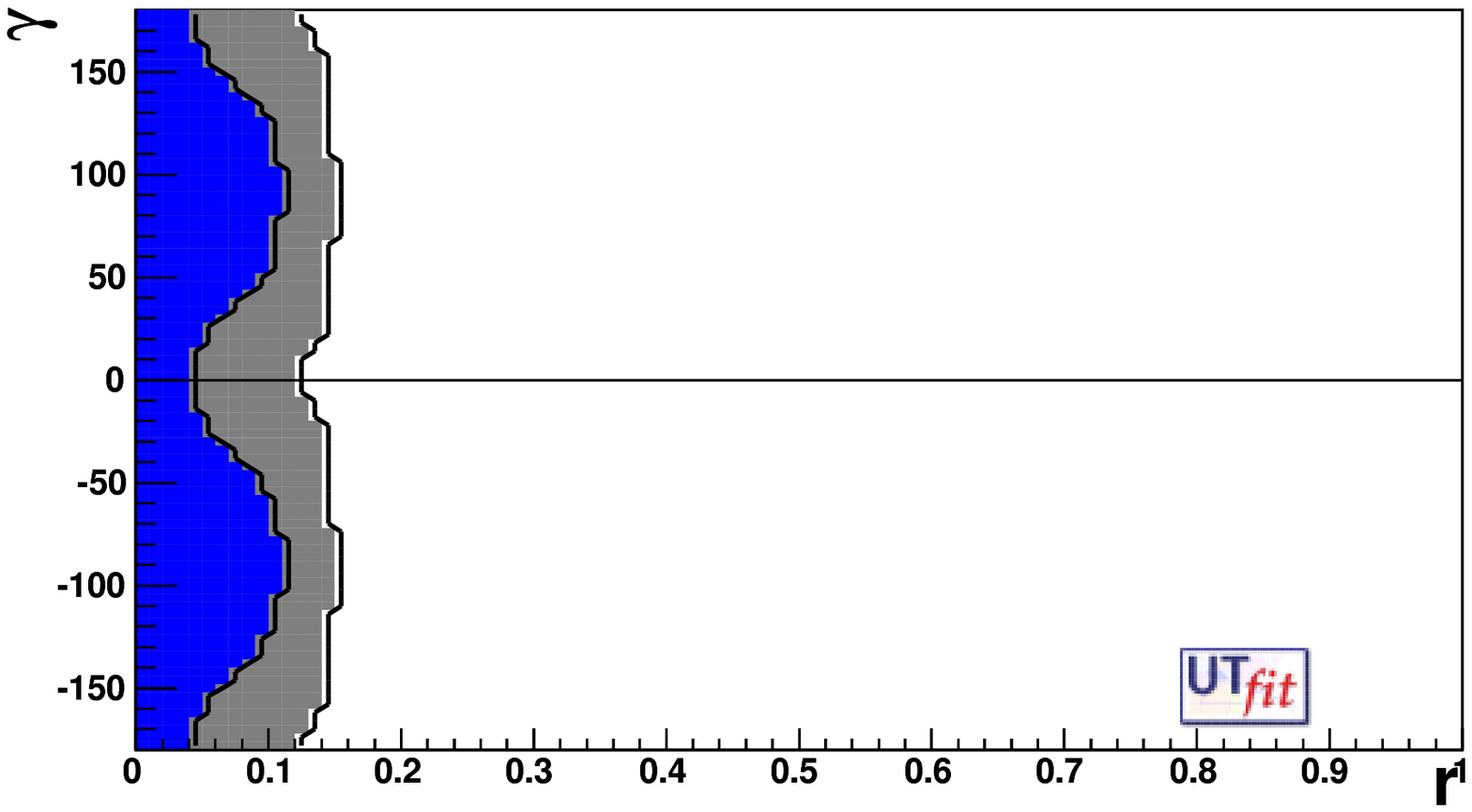}
\end{center}
\end{minipage}\hfill
\begin{minipage}{0.42\linewidth}
\begin{center}
\includegraphics*[width=1.\linewidth]{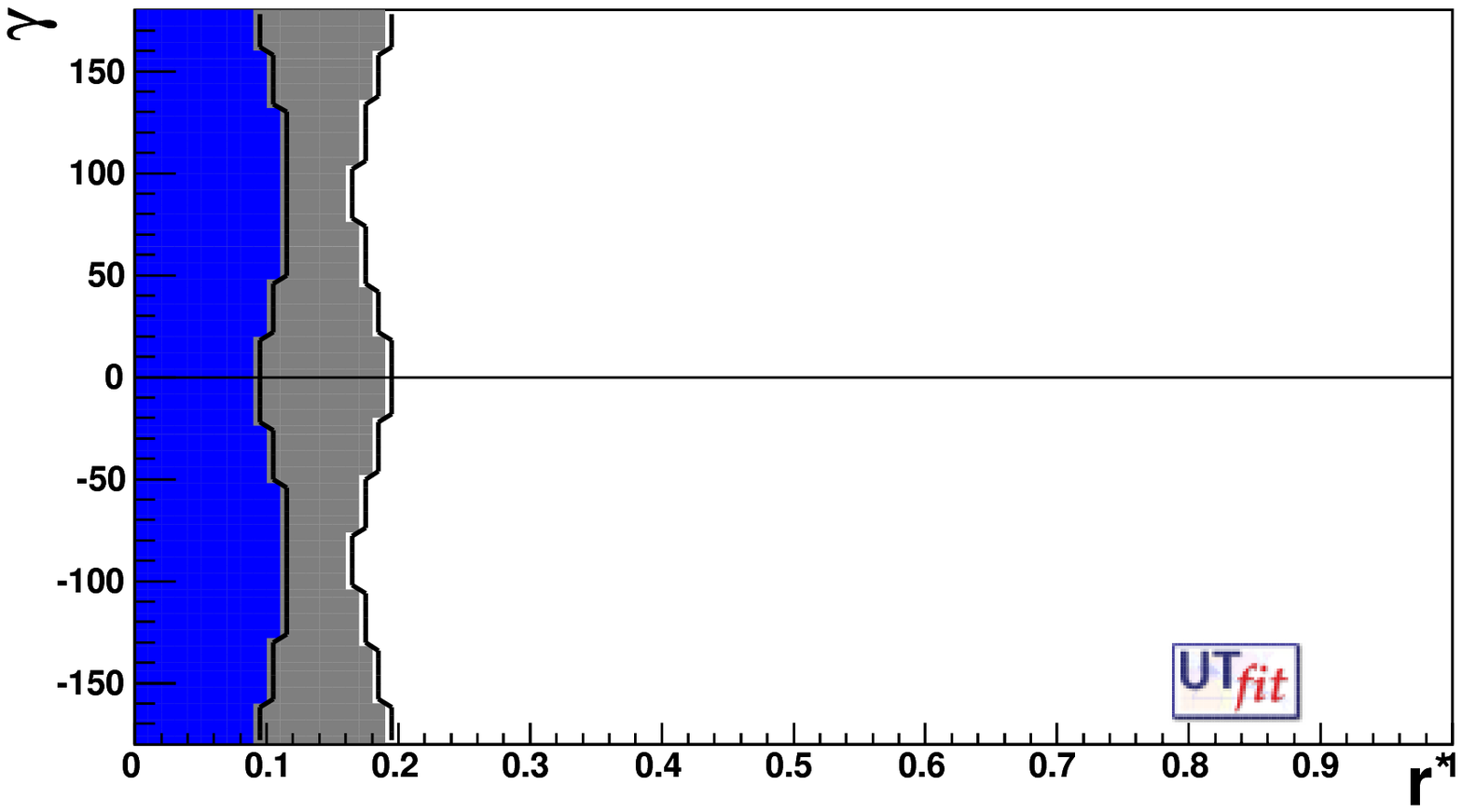}
\end{center}
\end{minipage}\hfill
\caption{68\% and 95\% C.L. as calculated by the UTfit group for $r_b$ for
the $D^{0}K$ mode (left) and for the $D^{*0}K$ mode (right) vs. $\gamma$. }
\label{utfit1}
\end{figure}

At \babar\ we have studied the $B^{\pm} \to D^0 K^{\pm}$ modes\cite{glw1} in the
flavor ($D^0\to K^-\pi+$), \CP$=+1$ ($D^0\to K^+K^-$ and 
$\pi^+\pi^-$), and \CP$=-1$ ($D^0\to K^0_S\pi^0$) eigenstates.
Figure\ref{8-0921_fig} shows the $B^-{\ra}D^0K^-$ signal
after requiring that the prompt track be consistent with the
kaon hypothesis for the flavor and \CP\ eigenstates. In a dataset of
$\sim$216 million $B\overline B$ pairs, we find 897 events in
the flavor,  93 in the \CP$=+1$, and 76 in the \CP$=-1$ eigenstates. 

The $B^{\pm} \to D^{*0} K^{\pm}$ modes\cite{glw2} have been studied, where the
$D^{*0}$ decays into $D^0\pi^0$, with the $D^0$ reconstructed in the
\CP\ even eigenstates $K^-K^+$ and $\pi^-\pi^+$, and in the flavor 
eigenstates  $K^-\pi^+$, $K^-\pi^+\pi^+\pi^-$, and $K^-\pi^+\pi^0$.
Figure~\ref{fit_con} shows the distributions of \deltae\ for the 
combined non-\CP\ and \CP\ modes before and after enhancing the $B\rightarrow
D^{*0}K$ component. This is accomplished by requiring that the prompt track be 
consistent with the
kaon hypothesis and that $\mes>5.27$ GeV/$c^2$. From a dataset of
$\sim$123 million $B\overline B$ pairs, we select 360  $B^{\pm} \to
D^{*0} K^{\pm}$ events in the non-\CP modes and 29 events in the \CP\
modes. The \deltae\
projections of the fit results are also shown.  

Finally we have studied the $B^{\pm} \to D^0 K^{*\pm}$ modes\cite{glw3} in the
flavor ($D^0\to K^-\pi+$, $K^-\pi^+\pi^+\pi^-$, and $K^-\pi^+\pi^0$),
\CP$=+1$ ($D^0\to K^+K^-$ and  
$\pi^+\pi^-$), and \CP$=-1$ ($D^0\to K^0_S\pi^0$, $K^0_S\omega$, $K^0_S\phi$)
eigenstates, with $K^{*-}\to K^0_S\pi^-$.
After requiring that the prompt track be consistent with the
kaon hypothesis, from a dataset of $\sim$227 million $B\overline B$
pairs, we find 498 events in flavor, 34 events in \CP$=+1$, and
15 events in \CP$=-1$ eigenstates.
\babar's results for the modes used in the GLW method are reported in
Table~\ref{table1}. 

\section{ADS related measurements}

We can also use the  Atwood, Dunietz and Soni
method, which exploits the interference between the decay chain
combining the CKM and color
suppressed $B^+\to D^0K^+$ decay and the CKM allowed $D^0\to K^-\pi^+$
decay and the one with a color allowed $B^+\to \overline{D^0}K^+$ decay
and the doubly CKM suppressed $\overline{D^0}\to K^-\pi^+$ decay. Using
this method we can measure: 
\begin{eqnarray}
R_{K\pi}  = \frac{\Gamma(B^- \to [K^+ \pi^-]_D K^-) + \Gamma(B^+ \to [K^-
\pi^+]_D K^+)}{\Gamma(B^- \to [K^- \pi^+]_D K^-) + \Gamma(B^+ \to [K^+ \pi^-]_D
K^+)} \nonumber\\ 
= r^2_b + r^2_d + 2r_d r_b \cos{\gamma}\cos{\delta}
\end{eqnarray}
where $r_d  =  |A(D^0 \to K^+ \pi^-)|/|A(D^0 \to K^- \pi^+)| = 0.060 \pm
0.003$, $\delta$ is the sum of the strong phase difference of the $B$s and
$D$s decay amplitudes and $r_d$ is 
the ratio of the suppressed $D$ decay to the dominant $D$ decay.

For this method we have studied both $B^{\mp} \to [K^{\pm} \pi^{\mp}]_D
K^{\mp}$ and $B^{\mp} \to [K^{\pm} \pi^{\mp}]_{D^*(D\pi/\gamma)}
K^{\mp}$ decays.
Figure~\ref{figads} shows the signal for the three suppressed decay modes.
\babar's results from a dataset of $\sim$227 million $B\overline B$ pairs
are consistent with no signal.
Using a Bayesian model, we measure: $r_b < 0.23$ at 90\%
C.L. for the $D^0K$ mode, and $r_b < 0.21$ at 90\%
C.L. for the $D^{*0}K$ mode\cite{D_Kpi_K}, as shown in
Figure~\ref{8-0948}, results which make a  
measurement of $\gamma$ quite difficult. 

\section{Constraints on $r_b$ and $\gamma$}

The latest conference results by \babar\ and Belle on the modes previoulsy
described have been combined by the UTfit\cite{utfit} group, and some of
the results, derived using a Bayesian approach, are reported in
Figure~\ref{utfit1}.

\section{Conclusions}

Many decays and methods have been or are being investigated to
extract the angle $\gamma$, and tighter constraints on its value will be
found, once larger data sets become available from both \babar\ and Belle,
though these measurements appear quite difficult given the latest \babar\
measurements of a small $r_b$ for the $D^0K$ and the $D^{*0}K$ modes.

\bibliographystyle{plain}

\end{document}